%% file: main.tex
\documentclass[submission,copyright,creativecommons]{eptcs}
\providecommand{\event}{ITRS 2016} 

\usepackage{etex}%to avoid problem when loading tikz package\underline{}

\usepackage{amssymb, amsmath,amsthm} %amssymb for \mathbb{} and \mathfrak{}, amsmath for \begin{equation*} and \varGamma
\usepackage{mathrsfs} %to use the style \mathscr{}
\usepackage{stmaryrd} %for \llbracket
\usepackage{cmll} %for symbols of linear logic
\usepackage{amsthm} %per non numerare i teoremi asteriscati
\usepackage{braket} %for the command \Set with scalable |
\usepackage{multirow} %for multicolumn and multirow spanning in tables
\usepackage{multicol} %for lists in multiple columns
\usepackage{color,soul} %per il testo barrato
\usepackage{enumitem} %per cambiare stile elenchi %to manage {itemize} and {enumerate}
\usepackage{microtype}
\usepackage{graphicx}
\usepackage{url}
\usepackage{wrapfig} %for floating figures
\usepackage{subcaption}
\usepackage[all]{xy}
\usepackage{hyperref}
\usepackage{tikz}
\usetikzlibrary{arrows}
\usetikzlibrary{automata}
\usepackage{pnGiulio}
\usepackage{ifthen}
\usepackage{adjustbox}
\usepackage{cleveref}

\hypersetup{unicode=true}
\input{Macros}

\newtheorem{theorem}{Theorem}

\newtheorem{corollary}[theorem]{Corollary}
\newtheorem{lemma}[theorem]{Lemma}

\theoremstyle{definition}

\newtheorem{definition}{Definition}

\title{\texorpdfstring{Relational type-checking for $\MELL$
    proof-structures.\\ Part 1: Multiplicatives}
  {Relational type-checking for MELL proof-structures. Part 1: Multiplicatives}}
\def\titlerunning{Relational type-checking for $\MELL$ proof-structures, 
  Part 1.}
\def\authorrunning{G.~Guerrieri, L.~Pellissier \& L.~Tortora de Falco}

\begin{document}

\author{ Giulio Guerrieri
  \institute{Dipartimento di Matematica e Fisica, Universit\`a Roma Tre, Rome,
    Italy}
  \institute{Institut de Math\'ematiques de Marseille, UMR 7373, Aix-Marseille
    Universit\'e, Centrale Marseille\\ F-13453 Marseille, France}
  \email{gguerrieri@uniroma.it}
  \and
  Luc Pellissier
  \institute{LIPN, UMR 7030, Universit\'e Paris 13, Sorbonne Paris Cit\'e\\
    F-93430 Villetaneuse, France}
  \email{luc.pellissier@lipn.univ-paris13.fr}
  \and
  Lorenzo Tortora de Falco
  \institute{Dipartimento di Matematica e Fisica, Universit\`a Roma Tre,
    Rome, Italy}
  \email{tortora@uniroma3.it}
}
\maketitle

\input{abstract}

\input{introduction}

\input{syntax}

\input{semantics}

\input{typing}

\input{cycles}

\input{vas}

\input{riam}

\input{recognition}

%% \input{future}

\input{lambda}

\section*{Acknowledgements}

Work partially supported by ANR projects C\textsc{oquas} ANR-12-JS02-006-01 and
E\textsc{lica} ANR-14-CE25-0005. The authors thank Damiano Mazza for pointing to
them the last Corollary.

\bibliographystyle{plain}
\bibliography{riam}
\addcontentsline{toc}{section}{References}%

%% \newpage
%% \appendix

%% \input{recognitionproof}

\end{document}

%% file: Macros.tex
\newcommand{\LL}{{\mathsf{LL}}}
\newcommand{\MLL}{\mathsf{M}\LL}
\newcommand{\ME}{\mathsf{ME}}
\newcommand{\MELL}{{\ME\LL}}

\newcommand{\Variables}{\mathcal{V}}
\newcommand{\Atom}{\mathcal{A}\!\mathit{t}}

\newcommand{\Cd}{\mathbin{\!\cdot\!}}

\newcommand{\Int}[1]{\lvert #1 \rvert_{\Variables}}

\newcommand{\Prin}[0]{\mathsf{pri}}
\newcommand{\Aux}[0]{\mathsf{aux}}

\newcommand{\Ports}[1]{\mathcal{P}(#1)}
\newcommand{\Cells}[1]{\mathcal{C}(#1)}
\newcommand{\Type}[1]{\mathsf{tp}_{#1}}
\newcommand{\PortsPrinFun}[1]{{\mathsf{P}^\Prin_{#1}}}
\newcommand{\PortsAuxFun}[1]{{\mathsf{P}^\Aux_{#1}}}

\newcommand{\Sem}[1]{\llbracket #1 \rrbracket}

\newcommand{\Nat}{{\mathbf{N}}}
\newcommand{\MultFin}[1]{\mathscr{M}_\mathrm{fin}(#1)}

\newcommand{\Mgu}{\mathrm{m.g.u.}}

\newcommand{\trans}{\xrightarrow}
\newcommand{\Environments}{\mathfrak{Env}}
\newcommand{\Bool}{\mathbf{B}}
\newcommand{\Trool}{\mathbb{B}_3}
\newcommand{\RelU}{\mathbb{REL}}
\newcommand{\Rel}{\mathbf{Rel}_{\oc}}
\newcommand{\Disp}{\mathbin{\delta}}

\newcommand{\reltype}[3]{\mathop{\rhd} {#1} : {#2} : {#3}}

%% file: abstract.tex
\begin{abstract}
  Relational semantics for linear logic is a form of non-idempotent intersection
  type system, from which several informations on the execution of a
  proof-structure can be recovered. An element of the relational interpretation
  of a proof-structure $R$ with conclusion $\Gamma$ acts thus as a type 
  (refining $\Gamma$) having $R$ as an inhabitant.

  We are interested in the following type-checking question: given a 
  proof-structure $R$, a list of formul\ae{} $\Gamma$, and a point $x$ in the 
  relational interpretation of $\Gamma$, is $x$ in the interpretation of $R$?
  This question is decidable. We present here an algorithm that decides it in 
  time linear in the size of $R$, if $R$ is a proof-structure in the 
  multiplicative fragment of linear logic. This algorithm can be extended to 
  larger fragments of multiplicative-exponential linear logic containing 
  $\lambda$-calculus.%Is it true? 
\end{abstract}

%% file: introduction.tex
\section{Introduction}
Intersection types have been introduced as a way of extending the
$\lambda$-calculus' simple types with finite polymorphism, by adding a new type
constructor $\cap$ and new typing rules governing it. A term of type $A \cap B$
can be used in further derivations both as data of type $A$ and as data of type
$B$. Contrarily to simple types (which are sound but incomplete), intersection
types present a sound and complete characterization of strong normalization.

Intersection types were first fomulated idempotent, that is, verifying the
equation $A \cap A = A$. This corresponds to an interpretation of a typed term
$M : A \cap B$ as \emph{$M$ can be used as data of type $A$ or as data of type
  $B$}. In a non-idempotent setting (\emph{i.e.}~by dropping the equation $A
\cap A = A$), the meaning of the typing judgment is strengthened to \emph{$M$
  can be used once as data of type $A$ and once as data of type $B$}.
Non-idempotent intersection types have been used to get qualitative and
quantitative information on the execution time of $\lambda$-terms
\cite{deCarvalho09,GrahamLengrand:2013df}.

Relational semantics is one of the simplest semantics of Linear Logic ($\LL$,
\cite{Girard87}). A $\LL$ formula is interpreted by a set, and a $\LL$
proof-structure \footnote{Following \cite{Girard87}, we make a difference
  between proof-structures and proof-nets: a proof-net is a proof-structure
  corresponding to a derivation in $\LL$ sequent calculus. Proof-nets can be
  characterized among proof-structures via “geometric” correctness criteria.}
by a relation between sets. Relational semantics correspond to a non-idempotent
intersection type system, called System $\mathsf{R}$ in \cite{deCarvalho09} (see
also \cite{PaoliniPiccoloRonchi15}).%% A type derivation in System $\mathsf{R}$
%% corresponds to an experiment; and the conclusion of such a type derivation is an
%% element of the relational semantics. All the qualitative and quantitative
%% information \cite{deCarvalho09,deCarvalhoPaganiTortora11} available through
%% non-idempotent intersection types can be recovered from the relational
%% semantics.

The relational semantics $\Rel$ of the $\lambda$-calculus arise from the
$\ast$-autonomous category of sets and relations and the co-monad of finite
multisets. Rather than describing its exact structure, we describe the
interpretation of simply-typed $\lambda$-terms, with base type $o$.

Let $\Atom$ be a set. To each type $\sigma$, we associate a set $\Sem{\sigma}$
as follows:
\begin{align*}
  \Sem{o} = \Atom \qquad \Sem{\sigma\to\tau} = \MultFin{\Sem{\sigma}} \times
  \Sem{\tau},
\end{align*}
where $\MultFin{\cdot}$ denotes the set of finite multisets. We will at times
write $X\to\alpha$ as a semantically-flavoured notation for pair $(X,\alpha) \in
\Sem{\sigma\to\tau}$.

To each valid typing sequent $\overline{x}:\overline{\sigma} \vdash M : \tau$
(where $\overline{x}:\overline{\sigma} = x_1 : \sigma_1, \dots, x_n :
\sigma_n$), we associate a set
\begin{align*}
  \Sem{\overline{x}:\overline{\sigma} \vdash M : \tau} \subseteq
  \MultFin{\Sem{\sigma_1}} \times \cdots \times \MultFin{\Sem{\sigma_n}} \times
  \Sem{\tau}
\end{align*}
as follows:
\begin{align*}
  \Sem{\overline{x}:\overline{\sigma} \vdash x_i : \sigma_i} &= \left\{
  (X_1, \dots, X_n, \alpha) \mid \alpha \in X_i \right\}\\  
  \Sem{\overline{x}:\overline{\sigma} \vdash \lambda y.M : \sigma \to
    \tau} &= \left\{ (X_1, \dots, X_n, Y \to \alpha) \mid (X_1, \dots, X_n, Y,
  \alpha) \in \Sem{\overline{x}:\overline{\sigma}, y : \sigma \vdash M :
    \tau} \right\}\\
  \Sem{\overline{x}:\overline{\sigma} \vdash M N : \tau} &= \left\{ (X_1, \dots,
  X_n, \alpha) \mid \exists Y \in \MultFin{\Sem{\sigma}}, (X_1, \dots, X_n, Y
  \to \alpha) \in \Sem{\overline{x}:\overline{\sigma} \vdash M : \sigma \to
    \tau} \right.\\
  &\qquad \forall \beta \in Y, (X_1, \dots, X_n, \beta) \in
  \Sem{\overline{x}:\overline{\sigma} \vdash N : \sigma} \}
\end{align*}
 As $\Rel$ is a cartesian closed category, if $M =_{\beta\eta} N$,
 $\Sem{\overline{x}:\overline{\sigma} \vdash M : \tau} =
 \Sem{\overline{x}:\overline{\sigma} \vdash N : \tau}$. We will write
 $\reltype{M}{\alpha}{\sigma}$ for $\alpha \in \Sem{\vdash M : \sigma}$,
 emphasizing that the intersection type $\alpha$ refines the simple type
 $\sigma$.

We now give examples of the kind of information that can be recovered from the
relational semantics:
\begin{itemize}
\item let $\Bool = o \to o \to o$, the Church encoding of booleans. Let
  $\mathbf{true} = \lambda x y. x$ and $\mathbf{false} = \lambda x y. y$.  Then
  we have $\reltype{\mathbf{true}}{[\ast] \to \varnothing \to \ast}{\Bool}$, but
  not $\reltype{\mathbf{false}}{[\ast] \to \varnothing \to \ast}{\Bool}$, where
  $\ast\in\Sem{o}$. As a consequence:
  \begin{theorem}
    \label{thm:boolean-semantic-evaluation}
    Let $M$ be a closed term of type $\Bool$. Then $M =_{\beta\eta}
    \mathbf{true}$ if and only if $\rhd M : [\ast] \to \varnothing \to \ast :
    \Bool$.
  \end{theorem}
  \item More generally, a $\lambda$-term $M$ has a sort of principal relational
    type: its 1-point, which can be computed efficiently when $M$ is in normal
    form. That is, given a term $M$ of type $\sigma$ in normal form, let
    $\mathsf{M}$ be its 1-point. Then, for any term $N$ of type $\sigma$, $M
    =_{\beta\eta} N$ if and only if $\reltype{N}{\mathsf{M}}{\sigma}$.
  \item Intersection types based on a variant of relational semantics have been
    shown useful \cite{GrelloisMellies15} to encode verification problems.
  \item A variant of relational semantics, Scott semantics, can be used as a
    faster alternative to $\beta$-evaluation \cite{Terui:2012go}.
  \item Given two terms $M_1 : \sigma \to \tau$ and $M_2 : \sigma$ in normal
    form, it is possible to compute the length of the reduction of $M_1M_2$ to
    its normal form \cite{deCarvalho09}.
\end{itemize}

%% \begin{itemize}
%% \item let $R$ be a proof-structure of conclusion $A$, and $x\in |A|$, where 
%% $|A|$ is the relational interpretation of $A$. If $S$ is another proof-structure
%% of type $A$, and $x$ is in the relational interpretation of $R$ but not $S$, 
%% then $R$ is not $\beta\eta$-equivalent to $S$. In particular, if $A$ is an 
%% encoding of the booleans, such a technique can be used to determine whether $R$
%% reduces to the encoding of $\mathbf{true}$ or the encoding of $\mathbf{false}$;
%% \item given two %correct (\emph{i.e.} arising from a
%% %   derivation on the sequent calculus) $\MELL$ proof-structure 
%%   $\LL$ proof-nets $\pi_1$ and $\pi_2$
%%   without cuts, it is possible to compute whether $\pi_1$ and $\pi_2$ can be
%%   composed and the length of the reduction to the normal form of this composition.
%% \end{itemize}

% Such information becomes valuable if it is easy to determine that a relational
% element is part (or not) of the relational interpretation of a
Such information becomes valuable when it is easy to determine whether a point
belongs to the relational interpretation of a proof-structure. In other word, we
are interested in the tractability of the following decision problem: given a
relational element $x$ and a $\lambda$-term $M$, can $M$ be typed by $x$ ?

As the simply-typed $\lambda$-calculus embeds in multiplicative-exponential
linear logic through the call-by-name translation $o \to o = \oc o \multimap o$,
we tackle this study on linear logic proof-nets.

As a first step towards the resolution of this question, we restrict ourselves
to proof-structures in the multiplicative fragment ($\MLL$) of $\LL$. In this 
particular setting, the interpretation of a formula is finite (up to innocuous 
renaming). We aim to climb in the ladder of several fragments of 
multiplicative-exponential $\LL$, providing algorithms of increasing complexity 
deciding this problem.

This problem has been present since the dawn of $\LL$; indeed, in its seminal
article, Girard \cite[3.16. Remark (ii), p. 57]{Girard87} answers the question
of the decidability of the following question: given a point $x$ and a proof
$\pi$, is $x$ in the coherent interpretation\footnote{The coherent semantics of
  linear logic being the one closest to the dynamics of
  cut-elimination. Relational semantics can be seen as a simplification of it.}
of $\pi$? The coherent setting is very different from the relational setting;
indeed, coherent semantics is not able to distinguish between certain
(non-connected) proof-structures \cite{Tortora03}. Nonetheless, we note along
with him that this problem – in the relational setting – is trivial for
multiplicative proof-structures without cuts: indeed, it suffices to propagate
the information present in the conclusions of a proof-structure. Cuts allow to
hide certain parts of the proof-structure from its conclusions (see
\Cref{fig:cyclic}); in the presence of cuts, cycles need a special treatment. In
this article, we will introduce a general framework deciding this problem for
multiplicative proof-structures with cuts, and explain how it can be adapted to
larger fragments of $\LL$ containing the $\lambda$-calculus.

We will define a variant of Vector Addition Systems (VAS, see
\cite{Leroux:2009br}) that encode naturally our decision problem. The machine
bears a close resemblance with the Interaction Abstract Machine (see for 
instance \cite{Laurent01}). It has indeed been known for a long time in the
Linear Logic community that Geometry of Interaction and relational semantics
enjoy a certain closeness. This work aims to bridge them on the operational 
side.

We provide an algorithm that decides in time linear in the size of the
multiplicative proof-structure $R$ whether a point $x$ is in the relational
interpretation of $R$. We give indications on how this algorithm can be extended
in a bilinear (in the size of the term and of the point) algorithm in the case
of $\lambda$-terms.

%% We introduce semantical typing judgments of the form $\vdash R : x : \parr
%% \Gamma$, where $R$ is a $\MELL$ (the multiplicative-exponential fragment of
%% linear logic) proof-structure whose conclusion is the list of $\MELL$
%% formul\ae{} $\Gamma$, and $x$ in the interpretation in the relational model of
%% the $\MELL$ formula $\parr\Gamma$.  Our goal is to decide in a tractable way
%% whether a judgment of this form is \emph{valid} or not, \textit{i.e.}~whether
%% $x$ is a point of the relational semantics of $R$ or not.

%% We thus define the \emph{Relational Interaction Abstract Machine} (Section
%% \ref{sec:RIAM}) able to decide such judgments on a fragment of all $\MELL$
%% proof-structures, that works by moving tokens embodying relational elements
%% through the proof-structure. The machine moreover stops on a sequent $\vdash R :
%% x : \parr \Gamma$ after a number of steps bilinear in the size of $x$ and of
%% $R$.

%% The class of $\MELL$ proof-structures on which our machine is sound and
%% complete, defined in Section \ref{sec:connection}, is moreover quite natural and
%% large enough to contain the $\lambda$-calculus.

%% file: syntax.tex
\section{\texorpdfstring{Elements of $\MLL$ syntax}
  {Elements of MLL syntax}}

%% We set $\mathcal{L}_{\MLL} = \{ 1, \bot, \otimes, \parr, \mathit{ax}, \mathit{cut}
%% \}$.  The \emph{$\MLL$ connectives } are $1, \bot, \otimes, \parr$.

The set of \emph{$\MLL$ formulas} is generated by the grammar:
\begin{equation*}
  A, B, C ::= X \ | \ X^\perp \ 
  | \ 1 \ | \ \bot \ 
  | \ A \otimes B \ | \ A \parr B \, .
\end{equation*}
where $X$ ranges over an infinite countable set of \emph{propositional
  variables}. %We extend $\cdot^{\bot}$ over all formul\ae{} as usual.
The linear negation $A^\perp$ of a formula $A$ is involutive, 
\textit{i.e.}~$A^{\perp\perp} = A$, and defined via De Morgan laws $1^\perp = 
\bot$ and $(A \otimes B)^\perp = A^\perp \parr B^\perp$.
If $\Gamma = (A_1,\dots , A_n)$ is a finite sequence of $\MLL$ formulas (with 
$n \in \Nat$), then $\parr\Gamma = A_1 \parr \dots \parr A_n$; in particular, 
if $n = 0$ then $\parr\Gamma = \bot$.

Proof-structures offer a syntax for a graphical representation of $\MLL$ proofs.
$\MLL$ proof-structures are directed labelled graphs $\Phi$ built from the
\emph{cells} defined in Figure~\ref{fig:cells}.
\input{figcells.tex}
We call \emph{ports} the directed edges of such graphs, labelled by $\MLL$
formulas. For every cell, its ports are divided into \emph{principal ports}
(outgoing in the cell, depicted down in the picture) and \emph{auxiliary ports}
(incoming in the cell, depicted up).

Let $\Phi$ be a $\MLL$ proof-structure. We denote by $\Ports{\Phi}$ the set of 
its ports and $\Cells{\Phi}$ the set of its cells. Let $c$ be a cell. We denote
by:
\begin{itemize}
\item $\Type{\Phi}(c)$ the type of $c$, ranging in $\{ 1, \bot, \otimes, \parr,
  \mathit{ax}, \mathit{cut} \}$;
\item $\PortsPrinFun{\Phi}(c)$ the principal ports of $c$. It is either
  \begin{itemize}
  \item a port, if $c$ is of type $1, \bot, \otimes, \parr$;
  \item an ordered pair of ports $\left\langle p_1,p_2 \right\rangle$, if $c$ is
    of type $\mathit{ax}$;
  \item empty, if $c$ is of type $\mathit{cut}$;
  \end{itemize}
\item $\PortsAuxFun{\Phi}(c)$ the auxiliary ports of $c$. It is either
  \begin{itemize}
  \item empty, if $c$ is of type $\mathit{ax}, 1, \bot$;
  \item an ordered pair of ports $\left\langle p_1, p_2 \right\rangle$, if $c$
    is of type $\mathit{cut}$, $\parr$ or $\bot$.
  \end{itemize}
\end{itemize}

% Among the ports of a $\MLL$-proof structure lie some that are only connected on
% one side. We call them the \emph{conclusions}. 
In a $\MLL$ proof-structure $\Phi$, any port that is principal for some cell but
is not auxiliary for any cell of $\Phi$ is called a \emph{conclusion} of $\Phi$.
We will only consider in the sequel $\MLL$ proof-structures with a fixed (total)
order on their conclusions. Given a list of $\MLL$ formul\ae{} $\Gamma=(A_1,
\dots,A_n)$ with $n \in \Nat$, we say that an $\MLL$ proof-structure $\Phi$ is
of \emph{conclusion} $\Gamma$ if the conclusions of $\Phi$ are the ordered
sequence $p_1 < \cdots < p_n$ of ports of $\Phi$ and $\Type{\Phi}(p_i)=A_i$ for
every $i \in \{1,\dots, n\}$.

The $\MLL$ proof-structure of \cref{fig:mll-ps} has two conclusions, one on the 
far right, the other on the far left. We will always depict proof-structures 
with conclusions ordered from left to right, so $R$ is of conclusion $(A\otimes 
B, A^{\bot}\parr B^{\bot})$.

%% file: figcells.tex
\begin{figure}[!t]
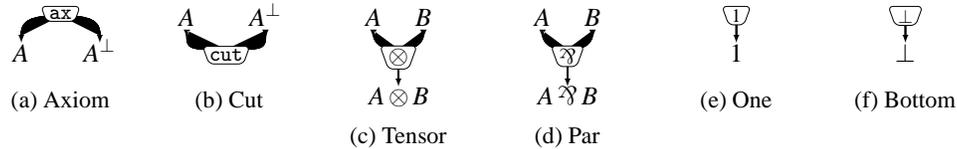

  \centering
  \vspace*{-3\baselineskip}
  \scalebox{0.9}{
  \begin{subfigure}[t]{0.15\textwidth}
    \centering
    \pnet{
      \pnformulae{
        \\
        \\
        \pnf[X]{$A$}~~\pnf[Xn]{$A^{\bot}$}
      }
      \pnaxiom[ax]{X,Xn}
    }
    \caption{Axiom}
  \end{subfigure}
  \begin{subfigure}[t]{0.15\textwidth}
    \centering
    \pnet{
      \pnformulae{
        \pnf[X]{$A$}~~\pnf[Xn]{$A^{\bot}$}
      }
      \pncut[cut]{X,Xn}
    }
    ~\\[-1cm]
    \caption{Cut}
  \end{subfigure}
  \begin{subfigure}[t]{0.15\textwidth}
    \centering
    \pnet{
      \pnformulae{
        \pnf[A]{$A$}~\pnf[B]{$B$}
      }
      \pntensor{A,B}[AoB]{$A\otimes B$}
    }
    \caption{Tensor}
  \end{subfigure}
  \begin{subfigure}[t]{0.15\textwidth}
    \centering
    \pnet{
      \pnformulae{
        \pnf[A]{$A$}~\pnf[B]{$B$}
      }
      \pnpar{A,B}[AoB]{$A\parr B$}
    }
    \caption{Par}
  \end{subfigure}
  \begin{subfigure}[t]{0.15\textwidth}
    \centering
    \pnet{
      \pnformulae{
        \\
        \\
        \pnf[A]{$1$}
      }
      \pninitial[1]{\scriptsize$1$}{A}
    }
    
    \caption{One}
  \end{subfigure}
  \begin{subfigure}[t]{0.15\textwidth}
    \centering
    \pnet{
      \pnformulae{
        \\
        \\
        \pnf[A]{$\bot$}
      }
      \pninitial[1]{\scriptsize$\bot$}{A}
    }
    \caption{Bottom}
  \end{subfigure}
  %% \begin{subfigure}[t]{0.15\textwidth}
  %%   \centering
  %%   \pnet{
  %%     \pnformulae{
  %%       \pnf[A1]{$A$}~\pnf[l]{$\cdots$}~\pnf[An]{$A$}
  %%     }
  %%     \pnexp{}{A1,l,An}[]{$\wn A$}
  %%   }
  %%   \caption{Why Not?}
  %% \end{subfigure}
  %% \begin{subfigure}[t]{0.15\textwidth}
  %%   \centering
  %%   \pnet{
  %%     \pnformulae{
  %%       \pnf[A1]{$A$}~\pnf[l]{$\cdots$}~\pnf[An]{$A$}
  %%     }
  %%     \pnbag{}{A1,l,An}[]{$\oc A$}
  %%   }
  %%   \caption{Of Course!}
  %% \end{subfigure}
  }
  \caption{The cells}
  \label{fig:cells}
\end{figure}

%% file: semantics.tex
\section{Elements of relational semantics}

We introduce here a variant of relational semantics (the simplest semantics of
Linear Logic, where formul\ae{} are interpreted by sets and proof-structures as
relations between sets) parametrized by a set $\Variables$ of variables. In this
variant, parts of a relational element can be left uninterpreted, allowing for
unification. We will use this feature in section \Cref{sec:cycles}.

\begin{definition}[Web of a $\MLL$ formula]
  Let $\Atom$ be a countably infinite set that doesn't contain the symbols of
  $\MLL$ or the empty sequence $(\,)$; the elements of $\Atom$ are called
  \emph{atoms}.

  Let $\Variables$ be a set disjoint from $\Atom$ whose elements are the
  \emph{atomic variables}.
  
  By induction, we define a function $\Int{\cdot}$ on $\MLL$ formul\ae{} by:
  \begin{align*}
    \Int{X^{\bot}} &= \Int{X} = \Atom \cup \Variables \text{, for all
      propositional variable } X\text{;}\\
    \Int{1} &= \Int{\bot} = \{()\};\\
    \Int{A \otimes B} &= \Int{A \parr B} = (\Int{A} \times \Int{B}) \cup
    \Variables,
  \end{align*}

  For a formula $A$, the set $\Int{A}$ is called the \emph{web of $A$}, whose
  elements are the \emph{points of $A$}.

  We write $\RelU=\bigcup_{A}\Int{A}$ the relational universe, where $A$
  range over all $\MLL$ formul\ae{}.
\end{definition}

The usual relational web of a $\MLL$ formula is recovered as
$|\Cd|_{\varnothing}$. We fix from now on an infinite set $\Variables$ of
variables.  Note that, for any $\MLL$ formula $A$, one has $\Int{A} =
\Int{A^\perp}$.
 
We define relational experiments straightforwardly on multiplicative
proof-structures by adapting the definition in \cite{Girard87}. Let $\rho$ be a
$\MLL$ proof-structure. A partial experiment of $\rho$ is a partial function of
the ports of $\rho$ associating with a port a relational element coherently with
the structure of $\rho$.

\begin{definition}[Experiment of a $\MLL$ proof-structure]
  Let $\Phi$ be a $\MLL$-ps.

  A \emph{partial experiment} $\mathsf{e}$ of $\Phi$ is a partial function
  associating with $p\in \Ports{\Phi}$ an element of $\Int{\mathsf{tp}_\Phi(p)}$
  verifying the following conditions, if $\mathsf{e}$ is defined on all the
  mentioned ports: let $c$ be a cell in $\Cells{\Phi}$,
  \begin{itemize}
  \item if $c$ is of type $\mathit{ax}$ with $\PortsPrinFun{\Phi}(c) =
    \left\langle p,q \right\rangle$, then $\mathsf{e}(p) = \mathsf{e}(q)$;
  \item if $c$ is of type $\mathit{cut}$ with $\PortsAuxFun{\Phi}(c) =
    \left\langle p,q \right\rangle$, then $\mathsf{e}(p) = \mathsf{e}(q)$;
  \item if $c$ is of type $1$ or $\bot$ with $\PortsPrinFun{\Phi}(c) =
    q$, then $\mathsf{e}(q) = ()$;
  \item if $c$ is of type $\otimes$ or $\parr$ with $\PortsAuxFun{\Phi}(l) =
    \left\langle p_1,p_2\right\rangle$, $\PortsPrinFun{\Phi}(l) = q$, then
    $\mathsf{e}(q) = (\mathsf{e}(p_1), \mathsf{e}(p_2))$.
  \end{itemize}

  An \emph{experiment} is a partial experiment defined on all ports, whose
  codomain can be restricted to $\bigcup_{p \in \Ports{\Phi}} 
  |\mathsf{tp}_\Phi(p)|_{\varnothing}$.
\end{definition}

If we consider the Cartesian product of sets and relations to be literally
associative, an experiment of a proof-structure of type $\Gamma=(A_1, \dots,
A_n)$ defines naturally an element of $\Int{\parr \Gamma}$, its
\emph{result}. We write $|\mathsf{e}|$ the result of an experiment
$\mathsf{e}$.

The relational interpretation of a $\MLL$ proof-structure $\Phi$ is then \(
\Sem{\Phi} = \{ |\mathsf{e}| : \mathsf{e} \text{ experiment of }\Phi\}.  \)

If we see the relational semantics as a non-idempotent intersection type system,
an experiment of a $\MLL$ proof-structure is a type derivation, and its result
is the conclusion of this type derivation. Just as we did for the
$\lambda$-calculus in the introduction, we write $\reltype{R}{\alpha}{\Gamma}$
for $\alpha \in \Sem{R} \subseteq \Int{\parr\Gamma}$. The point $\alpha$ acts
both as a witness of the fact that $\Gamma$ types $R$, while refining this type.

\begin{figure}[!th]
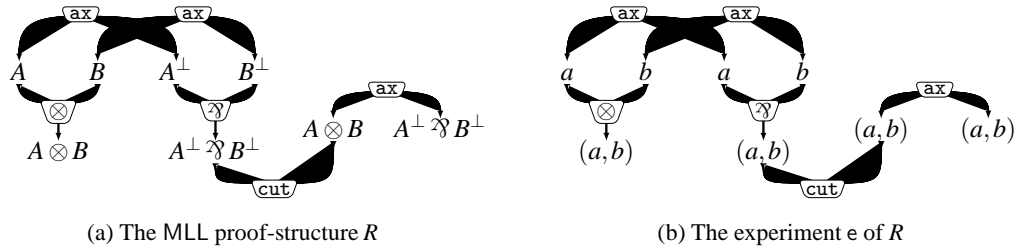

  \vspace*{-2\baselineskip}
  \centering
  \scalebox{0.9}{
    \begin{subfigure}[t]{0.5\textwidth}
      \centering
      \pnet{
        \pnformulae{
          \pnf[A1]{$A$}~~\pnf[B1]{$B$}~~\pnf[A2]{$A^{\bot}$}~~\pnf[B2]{$B^{\bot}$}\\
          ~~~~~~~~\pnf[AoB]{$A \otimes B$}~~~\pnf[ApB]{$A^{\bot} \parr B^{\bot}$}
        }
        \pnaxiom[axA]{A1,A2}[1.5][-0.25]
        \pnaxiom[axB]{B1,B2}[1.5][0.25]
        \pntensor{A1,B1}[tens]{$A \otimes B$}
        \pnpar{A2,B2}[par]{$A^{\bot}\parr B^{\bot}$}
        \pncut[cut]{par,AoB}
        \pnaxiom[axop]{AoB,ApB}
      }
      ~\\[-1cm]
      \caption{The $\MLL$ proof-structure $R$}
      \label{fig:mll-ps}
    \end{subfigure}
    \begin{subfigure}[t]{0.5\textwidth}
      \centering
      \pnet{
        \pnformulae{
          \pnf[A1]{$a$}~~\pnf[B1]{$b$}~~\pnf[A2]{$a$}~~\pnf[B2]{$b$}\\
          ~~~~~~~~\pnf[AoB]{$(a,b)$}~~~\pnf[ApB]{$(a,b)$}
        }
        \pnaxiom[axA]{A1,A2}[1.5][-0.25]
        \pnaxiom[axB]{B1,B2}[1.5][0.25]
        \pntensor{A1,B1}[tens]{$(a,b)$}
        \pnpar{A2,B2}[par]{$(a,b)$}
        \pncut[cut]{par,AoB}
        \pnaxiom[axop]{AoB,ApB}
      }
      ~\\[-1cm]
      \caption{The experiment $\mathsf{e}$ of $R$}
      \label{fig:mll-exp}
    \end{subfigure}
  }
  \caption{A $\MLL$-proof structure and an experiment on it}
  \label{fig:mll-ps-exp}
\end{figure}

The function depicted in \Cref{fig:mll-exp} is an experiment $\mathsf{e}$ of the
proof-structure of \Cref{fig:mll-ps}, where $a$ and $b$ are atoms. The
experiment $\mathsf{e}$ proves $\reltype{R}{((a,b),(a,b))}{(A \otimes B) \parr
  (A^{\bot} \parr B^{\bot})}$.

%% file: typing.tex
\section{Semantic typing}

%% Let $R$ be a $\MLL$ proof-structure. We introduce semantic typing judgments
%% of the form $\rhd R : x : \Gamma$, where $\Gamma$ is a list of formulas and
%% $x \in \lvert \parr\Gamma \rvert_{\Variables}$, meaning that
%% $x\in\Sem{R}$. That is, the point $x$ acts as a witness of the fact that
%% $\Gamma$ types $x$, while also giving information on the behaviour of $R$
%% under cut-elimination. Our contribution is to provide an algorithm deciding
%% this kind of judgments, with low complexity.

\begin{figure}[!ht]
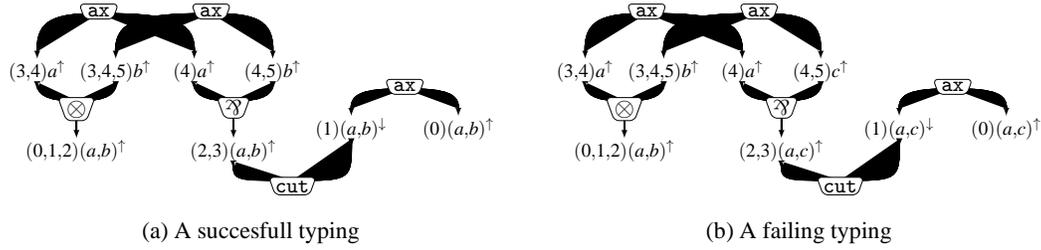

  \centering
  \scalebox{0.9}{
    \begin{subfigure}[t]{0.5\textwidth}
      \centering
      \pnet{
        \pnformulae{
          \pnf[A1]{$\scriptstyle(3,4) a^{\uparrow}$}~~\pnf[B1]{$\scriptstyle (3,4,5)
            b^{\uparrow}$}~~\pnf[A2]{$\scriptstyle (4) a^{\uparrow}$}
          ~~\pnf[B2]{$\scriptstyle (4,5) b^{\uparrow}$}\\
          ~~~~~~~~\pnf[AoB]{$\scriptstyle (1)
            (a,b)^{\downarrow}$}~~~\pnf[ApB]{$\scriptstyle (0) (a,b)^{\uparrow}$}
        }
        \pnaxiom[axA]{A1,A2}[1.5][-0.25]
        \pnaxiom[axB]{B1,B2}[1.5][0.25]
        \pntensor{A1,B1}[tens]{$\scriptstyle (0,1,2) (a,b)^{\uparrow}$}
        \pnpar{A2,B2}[par]{$\scriptstyle (2,3) (a,b)^{\uparrow}$}
        \pncut[cut]{par,AoB}
        \pnaxiom[axop]{AoB,ApB}
      }
      ~\\[-1cm]
      \caption{A succesfull typing}
      \label{fig:type-success}
    \end{subfigure}
    \begin{subfigure}[t]{0.5\textwidth}
      \centering
      \pnet{
        \pnformulae{
          \pnf[A1]{$\scriptstyle(3,4) a^{\uparrow}$}~~\pnf[B1]{$\scriptstyle (3,4,5)
            b^{\uparrow}$}~~\pnf[A2]{$\scriptstyle (4) a^{\uparrow}$}
          ~~\pnf[B2]{$\scriptstyle (4,5) c^{\uparrow}$}\\
          ~~~~~~~~\pnf[AoB]{$\scriptstyle (1)
            (a,c)^{\downarrow}$}~~~\pnf[ApB]{$\scriptstyle (0) (a,c)^{\uparrow}$}
        }
        \pnaxiom[axA]{A1,A2}[1.5][-0.25]
       \pnaxiom[axB]{B1,B2}[1.5][0.25]
        \pntensor{A1,B1}[tens]{$\scriptstyle (0,1,2) (a,b)^{\uparrow}$}
        \pnpar{A2,B2}[par]{$\scriptstyle (2,3) (a,c)^{\uparrow}$}
        \pncut[cut]{par,AoB}
        \pnaxiom[axop]{AoB,ApB}
      }
      ~\\[-1cm]
      \caption{A failing typing}
      \label{fig:type-fail}
    \end{subfigure}
  }

  \caption{Two typings or $R$}
  \label{fig:typing}
\end{figure}

We will now describe the main idea graphically, on two examples. In
\Cref{fig:typing}, we try, starting from the conclusions, to build an experiment
of $R$, one (\Cref{fig:type-success}) with putative result $((a,b),(a,b))$, the
other (\Cref{fig:type-fail}) with putative result $((a,b),(a,c))$ (with $b \neq
c$). Tokens travel through the proof-structure, encapsulating an element of the
relational interpretation of the port they are sitting on. We depict the tokens
at different step (where each step is defined by one token moving). Each token
is depicted as its travel direction, its content, and the step on which the
token exist: $(3,4) a^{\uparrow}$ meaning that a token is there on steps 3 and
4, containing $a$ and going up.

Let's describe possible execution steps in \Cref{fig:type-success}:
\begin{enumerate}
  \setcounter{enumi}{-1}
\item two upward tokens containing $(a,b)$ are placed on each conclusion 
  (the principal port of the $\otimes$-cell and the right principal port of the
  right axiom);
\item the right token goes up through the right axiom, and exits downwards from 
  its left principal port;
\item the same token goes down through the cut and exits upwards from its left 
  auxiliary port;
\item the other token (on the principal port of the $\otimes$-cell) gets split 
  in two upwards tokens, one containing $a$ on the left auxiliary port of the 
  $\otimes$-cell, the other containing $b$ on its right auxiliary port;
\item the right token containing $(a,b)$ on the principal port of the 
  $\parr$-cell gets split in two upwards tokens, one containing $a$, the other
  containing $b$;
\item the right token containing $a$ goes up through an axiom, exits downwards
  from its other principal port and meets an upwards token containing $a$
  too. They annihilate each other;
\item the right token containing $b$ goes up through an axiom, exits downwards
  from its other principal port and meets an upwards token containing $b$
  too. They annihilate each other.
\end{enumerate}
As there are no more tokens on the proof-structure, we say that the execution is
successful, and so we proved the judgment $\reltype{R}{((a,b),(a,b))}{(A \otimes
  B, A^{\bot}\parr B^{\bot})}$. Conversely, the same thing happens in
\Cref{fig:type-fail}, apart from the last step:
\begin{enumerate}
  \setcounter{enumi}{5}
\item the right token containing $b$ goes up through an axiom, exits downwards
  from its other principal port and meets an upwards token containing
  $c$. Nothing happens, the machine is stuck.
\end{enumerate}
As the machine is stuck with tokens on it, we say that the execution has failed,
and we proved the negation of the judgment $\reltype{R}{((a,b),(a,c))}{(A \otimes
  B, A^{\bot}\parr B^{\bot})}$.

We will formalize this mechanism in \Cref{sec:RIAM}.

%% file: cycles.tex
%% \section{Accommodating ``semantically hidden'' cycles}

\label{sec:cycles}

\begin{wrapfigure}[9]{l}{0.4\textwidth}
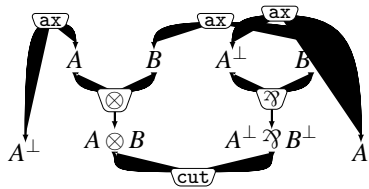

  ~\\[-1cm]
  \centering
  \scalebox{0.9}{
    \pnet{
      \pnformulae{
        ~\pnf[A1]{$A$}~~\pnf[B1]{$B$}~~\pnf[A2]{$A^{\bot}$}~~\pnf[B2]{$B^{\bot}$}\\
        \\
        \pnf[A'1]{$A^{\bot}$}~~~~~~~~\pnf[A'2]{$A$}
      }
      \pnaxiom[axA1]{A1,A'1}
      \pnaxiom[axB]{B1,B2}[1][-0.25]
      \pnaxiom[axA2]{A2,A'2}[1.2][-0.25]
      \pntensor{A1,B1}[tens]{$A \otimes B$}
      \pnpar{A2,B2}[par]{$A^{\bot}\parr B^{\bot}$}
      \pncut[cut]{par,tens}
    }
    }
  ~\\[-1cm]
  \caption{A cyclic proof-structure $R$}
  \label{fig:cyclic}
\end{wrapfigure}

We want our algorithm to be able to handle cases like the proof-structure in
\Cref{fig:cyclic}, where part of the information carried by an experiment can
not be retrieved from its conclusion.

By simply propagating a relational point using the same strategy as defined
before, we cannot guess which relational element should be the image of the port
of type $B$ through the experiment. In a way, all the information in the cycle
concerning $B$ is hidden from the conclusions. We solve this problem by
introducing variables (which we already forced into the definition of the
relational web of a formula): some transitions can be fired with incomplete
information which will be checked later.

%%  We give a sequence of transitions giving a proof
%% of $\vdash R : (a,a) : (A^{\bot},A)$, depicted in \Cref{fig:cyclic-run}:
%% \begin{enumerate}
%%   \setcounter{enumi}{-1}
%% \item two upward tokens containing $a$ are placed on each conclusion (the left 
%%   principal port of the left axiom and the right principal port of the right 
%%   axiom);
%% \item the left upward token goes up through the left axiom;
%% \item the right upward token goes up through the right axiom;
%% \item the right token goes down through the $\parr$-cell, creating two tokens 
%%   and a fresh variable $\bullet$ on the fly;
%% \item the left token goes down through the $\otimes$-cell, creating two tokens 
%%   and a fresh variable $\circ$ on the fly;
%% \item the token containing $\bullet$ goes up through the axiom and meets a token
%%   containing $\circ$. As $\bullet$ is a variable, it gets unified with $\circ$,
%%   the environment of the machine stores $\bullet\mapsto\circ$ and the two tokens
%%   annihilate each other;
%% \item the token containing $(a,\circ)$ goes down through the cut. As $(a,\circ)$
%%   and $(a,\bullet)$ are unifiable (they are already unified) under the
%%   environment $\bullet\mapsto\circ$, they annihilate each other.
%% \end{enumerate}
%% The machine stops with no more tokens; we proved that $\vdash R : (a,a) : (A^{\bot},A)$.

%% file: vas.tex
\section{Vector Addition Systems}

\label{sec:vas}

Vector Addition Systems (VAS) (for instance, \cite{Leroux:2009br}) are a
prominent class of infinite state systems. They comprise a finite number of
counters ranging over the (non-negative) natural numbers. When taking a
transition, an integer can be added to a counter, provided it stays positive. To
give an example, let us consider a VAS with two counters. Let $\alpha$ be
defined as the displacement $(1,-1)$. The transition $(1,2)\trans{\alpha}(2,1)$
is valid, while $\alpha$ does not define any transition starting from the state
$(1,0)$, because $0+(-1)$ is negative.

VAS are particularly well-suited to represent systems with an infinite number
of states. Our idea here is to encode the presence, direction and content of a
token in a variant of VAS: to each port $p$ is associated a counter, which is
set to 0 if there is no token on $p$, $a$ if there is an upwards token
containing the relational element $a$ on $p$, $-a$ if there is a downwards token
containing $a$ on $p$. While the systems studied in the sequel have an
essentially finitary behaviour, it is not the case in extensions to the
exponentials of linear logic: the relational type-checking of $\MELL$-proof
structures with exponentials will imply the presence of any arbitrary number of
tokens on a given port.

All counters in VAS are natural numbers. We depart from traditional VASs for two
reasons: we want to be able to encode tokens going up, but also going down. We
also want counters to account for tokens containing a relational element, so the
counters are to be taken is a space engendered by relational elements with
coefficients. For multiplicative proof-structures, we can restrict ourselves to
the case where every coefficient is in $\Trool = \{-1,0,1\}$. As such, the
counter associated to a port $p \in \Ports{\Phi}$ is a formal series with
coefficients in $\{-1, +1\}$. We denote by $\Trool[\Int{\Type{\Phi}(p)}]$ the 
set of such formal series. It is endowed with a partial sum and a partial 
difference (by extending pointwise the partial sum and partial difference of 
$\Trool$, seen as a subset of $\mathbf{Z}$). The configuration of the machine 
ought then to be an element of the dependent product $\prod_{p \in \Ports{\Phi}}
\Trool[\Int{\Type{\Phi}(p)}]$, which we will write as a function.

%% file: riam.tex
\section{The relational interaction abstract machine}

\label{sec:RIAM}

We are now ready to give the formal definition of the Relational Interaction
Abstract Machine for $\MLL$, which decides semantic typing judgments.

Given a relation $R\subseteq A\times B$ and any $a\in A $ and $ b\in B$, $a \, R
\, b$ stands for $(a,b)\in R$.

\begin{definition}[Relational Interaction Abstract Machine for $\MLL$]
  Let $\Phi$ be a $\MLL$ proof-structure.

  An \emph{environment} is a finite map from $\Variables$ to $\RelU$. The set
  of environments of $\Phi$ is denoted by $\Environments$.
  
  The \emph{relational interaction abstract machine for $\MLL$} (RIAM)
  associated with $\Phi$, denoted by $M^{\Phi}$, has the following components:
  \begin{itemize}
  \item its alphabet is $\Sigma=\Environments \cup \Cells{\Phi}$, where
    $\Cells{\Phi}$ is the set of cells of $\Phi$;
  \item its set of configurations is $\prod_{p \in \Ports{\Phi}}
    \Trool[\Int{\Type{\Phi}(p)}]$.
  \end{itemize}
  The RIAM associated with $\Phi$ has two kinds of transitions: the
  \emph{displacement transitions} or the \emph{unification transitions}.

  The displacement transitions are labelled by an element $c$ of
  $\Cells{\Phi}$. The binary relation $\trans{c}$ on configurations of
  $M^{\Phi}$ is defined by:
  \[
  x \trans{c} x'\ \text{ if } c \Disp (x-x').
  \]
  where the \emph{displacement relation} $\delta \subseteq \Cells{\Phi} \times
  \prod_{p \in \Ports{\Phi}} \Trool[\Int{\Type{\Phi}(p)}]$ is the relation
  defined by:
  \begin{itemize}
  \item if $c$ is of type $\mathit{ax}$, let $\PortsPrinFun{\Phi} (c) =
    \left\langle p,q\right\rangle$ and
    $
    \forall a \in\Int{\Type{\Phi}(p)},
    c \Disp \left\{\begin{array}{rcl}
    p&\mapsto&-a\\
    q&\mapsto&-a\\
    r&\mapsto& 0\text{, if }r\neq p,q
    \end{array} \right.
    $
  \item if $c$ is of type $\mathit{cut}$, let $\PortsAuxFun{\Phi} (c) =
    \left\langle p,q\right\rangle$ and
    $
    \forall a \in \Int{\Type{\Phi}(p)},
    c \Disp \left\{\begin{array}{rcl}
    p&\mapsto&a\\
    q&\mapsto&a\\
    r&\mapsto& 0\text{, if }r\neq p,q
    \end{array} \right.
    $
  \item if $c$ is of type $1$ or $\bot$, let $\PortsPrinFun{\Phi} (c) = p$ and
    $
    c \Disp \left\{\begin{array}{rcl}
    p&\mapsto&-(\,)\\
    r&\mapsto& 0\text{, if }r\neq p
    \end{array} \right.
    $
  \item if $c$ is of type $\otimes$ or $\parr$, let $\PortsPrinFun{\Phi}= q$
    and $\PortsAuxFun{\Phi} = \left\langle p_1, p_2\right\rangle$, and
    $
    c \Disp \left\{\begin{array}{rcl}
    p_1&\mapsto& \bullet\\
    p_2&\mapsto& \circ\\
    q&\mapsto& -(\bullet, \circ)\\
    r&\mapsto& 0\text{, if }r\neq p_1,p_2,q
    \end{array} \right..
    $
    where $\bullet,\circ \in \Variables$ are two fresh variables.
  \end{itemize}

  The unification transitions are labelled by environments and defined on
  couples of configurations by:
  \[
  x \trans[\mathbf{u}]{s} s(x) \text{ if } \exists p\in\Ports{\Phi}, \left\{
  \begin{array}{l}
    x(p) = a_1 - a_2 + \vec{a}, \ a_1\neq a_2, \ \vec{a}\in
    \Trool[\Int{\Type{\Phi}(p)}]\\
    s = \Mgu(a_1,a_2)
  \end{array}\right.
  \]
  where $\Mgu$ denotes the most general unifier. Such a transition unifies (and
  so annihilates) two elements of a formal sum of opposite sign in one of the
  counters.

  We define $\trans[]{\sigma}$, for $\sigma \in \Sigma^{\star}$ by relational
  composition: let $\sigma=a_1s_2\cdots a_{n-1}s_n$, we set
  \[
  x \trans[]{\sigma} x' \text{ if } \exists (x_i)_{1 \leqslant i < n}, x
  \trans[]{a_1} x_1 \trans[\mathbf{u}]{s_2} \cdots \trans[]{a_{n-1}} x_{n-1}
  \trans[\mathbf{u}]{s_n} x.
  \]
  We say that $\sigma$ is an \emph{execution} of $M^{\Phi}$. 

  We denote by $\trans[\Phi]{}$ the \emph{reachability binary relation}
  defined by
  $
  x \trans[\Phi]{} x' \text{ if } \exists \sigma \in \Sigma^{\star}, x
  \trans[]{\sigma} x'.
  $

  We say that an element $x\in \prod_{l \in \Ports{\Phi}}
  \Trool[\Int{\Type{\Phi}(l)}]$ is \emph{accepted} or \emph{recognized} by
  $M^{\Phi}$ if
  $
  x \trans[\Phi]{} 0.
  $
  and that a $\sigma\in\Sigma^{\star}$ accepts (or recognizes) $x$ if $x
  \trans[]{\sigma} 0$.

  If $x=(x_1,\dots, x_n) \in \Int{\parr\Gamma}$, where $\Gamma= (A_1,
  \dots,A_n)$ and the $\MLL$ proof-structure $\Phi$ has $n$ conclusions $p_1 <
  p_2 < \cdots < p_n$, we say that $M^{\Phi}$ \emph{accepts} $x$ if it accepts
  the element of $\prod_{p \in \Ports{\Phi}} \Trool[\Int{\Type{\Phi}(p)}]$
  associating $0$ with every port which is not a conclusion, and $x_i$ to $p_i$.
\end{definition}

\begin{wrapfigure}[7]{r}{0.4\textwidth}
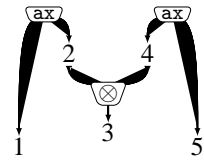

  ~\\[-1.6cm]
  \centering
  \scalebox{0.9}{
    \pnet{
      \pnformulae{
        ~\pnf[A1]{2}~~\pnf[B1]{4}\\
        \\
        \pnf[A'1]{1}~~~~\pnf[A'2]{5}
      }
      \pnaxiom[axA1]{A1,A'1}
      \pnaxiom[axB]{B1,A'2}[1][0]
      \pntensor{A1,B1}[tens]{3}
  }}
  \caption{A named proof-structure $\Phi$}
  \label{fig:cyclic-ports}
\end{wrapfigure}

%% \begin{example}
We will describe the RIAM associated to the proof-structure $\Phi$ in
\Cref{fig:cyclic-ports}. Its ports are numbered, its cells are named $ \{
\texttt{ax}_{1, 2}, \texttt{ax}_{4,5}, {\scriptstyle \otimes} \}$.

The RIAM associated to $\Phi$ has $\Sigma = \Environments \cup \{
\texttt{ax}_{1, 2}, \texttt{ax}_{4,5}, {\scriptstyle \otimes} \}$ as alphabet,
$\Int{A} \times \Int{A} \times \Int{A \otimes B} \times \Int{B} \times \Int{B}$
as set of configurations and its displacement relation $\delta$ is defined by,
for $\circ, \bullet$ fresh variables:
\begin{align*}
  \forall a \in \Int{A}, \texttt{ax}_{1,2} &\mathbin{\delta} (-a, -a, 0, 0, 0)\\
  \forall b \in \Int{B}, \texttt{ax}_{4,5} &\mathbin{\delta} (0, 0, 0, -a, -a)\\
  {\scriptstyle \otimes} &\mathbin{\delta} (0, \circ, -(\circ, \bullet), \bullet, 0).\\
\end{align*}
The relational element $(a,(a,b),b)$ is recognized by the word $\texttt{ax}_{1,
  2} {\scriptstyle \otimes} \{\bullet \mapsto b, \circ \mapsto a\}
\texttt{ax}_{4, 5}$. %% as witnessed by the sequence of states:
%% \begin{align*}
%%   (a, 0, (a,b), 0, b)
%%   &\trans{\texttt{ax}_{1, 2}} (0, -a, (a,b), 0, b)\\
%%   &\trans{\otimes} (0, \circ - a, (a,b)-(\circ,\bullet), \bullet, b)\\
%%   &\trans[\mathbf{u}]{\bullet \mapsto b, \circ \mapsto a} (0, 0, 0, b, b)\\
%%   &\trans{\texttt{ax}_{4, 5}} (0, 0, 0, 0, 0)
%% \end{align*}
%% %% \end{example}

%% file: recognition.tex
\section{Recognition of the relational interpretation}

\label{sec:recognition}
Among all executions of a RIAM, some are in normal form: intuitively, they don't
create any tokens but only propagate those already present in the initial
configuration.%%  Normal execution on initial configurations derived from
%% relational points are instrumental in the sequel, as they have good structure
%% (see \Cref{lem:normal-are-good}) and successful executions on such
%% configurations can be normalized (see \Cref{lem:normalization}).

\begin{definition}[Normal execution]
  Let $\Phi$ be a $\MLL$ proof-structure. An execution $\sigma$ of $M^{\Phi}$ is
  \emph{normal} if
  \begin{itemize}
  \item for each displacement transition $x\trans{c}x'$ in $\sigma$, there
    exists a $p\in\Ports{\Phi}$ such that $x(p)\neq 0$ and $x'(p)=0$;
  \item each displacement transition $x\trans{c}x'$ in $\sigma$ such that there
    exists $p \in \Ports{\Phi}$ such that $x'(p)=-a+a'$ is followed by an
    unification transition $x'\trans[\mathbf{u}]{s}x''$ such that $x''(p) = 0$.
  \end{itemize}
\end{definition}

\begin{lemma}
  \label{lem:normal-are-good}
  Let $\Phi$ be a $\MLL$ proof-structure of conclusion $\Gamma$.
  Let $x \in \lvert \parr\Gamma \rvert_{\varnothing}$.

  %% A normal successful run $\sigma$ of $M^{\Phi}$ on $x$ is of the form
  %% $\sigma=\sigma_1 \cdots \sigma_m$, where each $\sigma_i$ is a succession of
  %% displacement transitions followed by a unification transition (except
  %% eventually $\sigma_m$).

  %% Moreover, for each configuration $x$ attained after any $\sigma_i$, for all $p
  %% \in \Ports{\Phi}$, $x(p)=0$ or there exists $a\in\RelU$ such that $x(p) = \pm
  %% a$.

  %% Finally, $\sigma$ is at most of length $2 |\Cells{\Phi}|$, where 
  %% $|\Cells{\Phi}|$ is the number of cells of $\Phi$.

  A normal successful run of $M^{\Phi}$ on $x$ is at most of length twice the
  number of cells of $\Phi$.
\end{lemma}

\begin{lemma}
  \label{lem:normalization}
  Let $\Phi$ be a $\MLL$ proof-structure with conclusion $\Gamma$.
  Let $x \in \lvert\parr\Gamma\rvert_{\varnothing}$ be such that $M^{\Phi}$ 
  accepts $x$.
  Then, there exists a normal execution of $M^{\Phi}$ that recognizes $x$.
\end{lemma}

Normal executions of the machine can be used to define a partial experiment on
all ports connected to the conclusions, while an experiment can be
sequentialized in an execution.

\begin{theorem}
  \label{thm:main}
  Let $\Phi$ be a $\MLL$ proof-structure with conclusions $\Gamma$ (a list of
  $\MLL$ formul\ae{}). Let $x\in \lvert\parr\Gamma\rvert_{\varnothing}$.

  %The judgment $\vdash \Phi : x : \Gamma$ is valid
  
  Then, $x \in \Sem{\Phi}$ if and only if there exists a (normal) execution of
  $M^{\Phi}$ recognizing $x$.

  %% Moreover, if there is an execution of $M^{\Phi}$ recognizing $x$ then all
  %% normal executions of $M^{\Phi}$ recognizing $x$ (and there exists at least 
  %% one) have length $O(|\Cells{\Phi}|)$.
\end{theorem}

As moreover, a run with a counter in a state of the form $a+b$, with $a$ and $b$
not unifiable cannot be extended in a successful run, we get:

\begin{corollary}
  The RIAM decides judgements of the form $\rhd \Phi : x : \Gamma$ in time
  $O(|\Cells{\Phi}|)$ (\textit{i.e.}~in time linear in the size of $\Phi$) by
  attempting a normal run which sequence of displacement transition refine the tree
  order of cells in $\Phi$.
\end{corollary}

%% In other words, given a $\MLL$ proof-structure $\Phi$ with conclusion the list 
%% of $\MLL$ formul\ae{} $\Gamma$ and given $x\in 
%% \lvert\parr\Gamma\rvert_{\varnothing}$, the RIAM $M^{\Phi}$ decides judgments of
%% the form $\rhd \Phi : x : \Gamma$ (\textit{i.e.}~if $x \in \Sem{\Phi}$ or $x 
%% \notin \Sem{\Phi}$) in time $O(|\Cells{\Phi}|)$, \textit{i.e.}~in time linear in
%% the size of $\Phi$.

%% file: lambda.tex
\section{Extending to the lambda-calculus}

The techniques in this article can be extended to the $\lambda$-calculus
translated in $\MELL$ proof-structure through the call-by-name translation
$\alpha \to \beta = \oc \alpha \multimap \beta$. The formul\ae{} of $\MELL$ are
those of $\MLL$ to which is added two new connectives, $\oc$ and $\wn$. The
syntax of the proof-structure is enriched with a $\oc$-cell, having an arbitrary
number of unordered inputs, and a box constructor, taking a proof-structure and
encapsulating.

The interpretation of the exponential cells are a bit tedious to define. An
elegant way is described in \cite{GuerrieriPellissierTortora16}. It is multi-set
based: the interpretation of a formula $\oc A$ or $\wn A$ is a multiset. The
interpretation of a box is the multiset containing multiple interpretations of
the content of the box.

We extend the definition of the abstract machine to handle the exponentials:
\begin{itemize}
\item coefficients of the machine have to be taken in $\mathbf{Z}$ (and no more
  in $\Trool$), moreover on top of containing a relational element, tokens
  contain also a stack of timestamps remembering when boxes are entered;
\item new rules must be added, allowing to pass through exponential cells. They
  amount to splitting the content of multi-cells in all possible ways;
%% \item a new definition of normal run must be used. Several are possible, the one
%%   adapted to the $\lambda$-calculus consists of only passing through
%%   $\oc$-cells.
\end{itemize}
Together with an appropriate definition of normal run, this allows to prove:
\begin{theorem}
  The RIAM for the $\lambda$-calculus decides judgements of the form
  $\reltype{M}{x}{\Gamma}$ in time $O(|M|\times |x|)$.
\end{theorem}

As a corollary of this Theorem and of \Cref{thm:boolean-semantic-evaluation}, we
get the following (unpublished) result:
\begin{corollary}[Terui, 2012]
  Let $\mathbf{W} := (o \to o) \to (o \to o) \to o \to o$ be the Church encoding
  of binary strings.

  Let $M$ be a closed $\lambda$-term of type $\mathbf{W}\to\mathbf{B}$. It
  decides a language $\mathcal{L}$.

  $\mathcal{L}$ is in $\mathbf{LinTIME}$ (deterministic linear time).
\end{corollary}
The proof consisting of checking, for an encoded string $s : \mathbf{W}$,
whether $\rhd Ms : [\ast] \to \varnothing \to \ast : \mathbf{B}$, which is done
in time linear in the size of the translation of $M$, itself linear in the size
of $M$.

The result is surprising, as simply-typed $\lambda$-terms of type $\mathbf{N}
\to \mathbf{N}$ (where $\mathbf{N}$ is the Church encoding of natural numbers)
can represent a function of complexity an arbitrary tower of exponentials.